# Practical Metal-Wire THz Waveguides

Andrey Markov, Stephan Gorgutsa, Hang Qu and Maksim Skorobogatiy[*]

*Department of Engineering Physics, École Polytechnique de Montréal, Québec, Canada*
*Corresponding author: maksim.skorobogatiy@polymtl.ca

Several practical implementations of the THz wire waveguides are proposed combining the low-loss, low group velocity dispersion and simplicity of excitation of the two-wire metal waveguide with the ease of manufacturing, possibility of bending and a convenient access to a modal power of the hollow core waveguide. Optical properties of the wire waveguides are investigated experimentally by THz-TDS spectroscopy and numerically using the finite element method confirming TEM-mode guidance in a broad spectral range.

The main complexity in designing terahertz waveguides is the fact that almost all materials are highly absorbent in the terahertz region [1]. Since the lowest absorption loss occurs in dry air, an efficient waveguide design must maximize the fraction of power guided in the air. Different types of THz waveguides and fibers have been proposed based on this concept. The simplest subwavelength fiber [2-4], features dielectric core that is much smaller than the wavelength of guided light. As a result, a high fraction of modal power is guided outside of the lossy material and in the low-loss gaseous cladding. Another type of low-loss fibers features porous core region with the size of the individual pores much smaller than the wavelength of light [2, 3]. As a result, guided light is concentrated mostly in the low-loss gas-filled pores inside the core and in the gaseous cladding. Porous fibers, generally, feature higher confinement in the core region and are less prone to bending losses and influence of the environment compared to simple rod-in-the-air subwavelength fibers [2, 7]. Another important type of the low-loss waveguides feature hollow gas-filled core surrounded by a structured cladding serving as a reflector. The main challenge in the design of such waveguides is to ensure high reflection at the core-cladding interface. Different hollow-core structures have been investigated including metalized bores [5-7], periodic dielectric multilayers [8], as well as thin-walled dielectric pipes [9-11]. Another types of the THz waveguide are parallel plate waveguides [12] and slit waveguides [13] remarkable for their low transmission losses and high mode confinement.

Metal wire, can be used to transport terahertz pulses with virtually no dispersion and low attenuation [14.]. However, it is difficult to couple an electromagnetic wave efficiently to the radially-polarized mode supported by this waveguide. The primary mode of a single wire is radially polarized, thus, commonly used linearly polarized THz antennas cannot be used for the efficient excitation of this mode. Also high losses when the wire is bent limits the practical applications of the single wire waveguide.

Two-wire waveguides [15] combine both low loss and efficient coupling properties. The modal pattern of this type of waveguide (see Fig. 1) is very similar to the field emitted from a simple dipole, resulting in efficient coupling of the linearly polarized electromagnetic field from typical terahertz sources into that mode. Also, the group velocity dispersion of the TEM mode of this waveguide is extremely low (see Fig. 2). The confinement of the modal power in a small area between the metal wires opens possibilities for various practical applications.

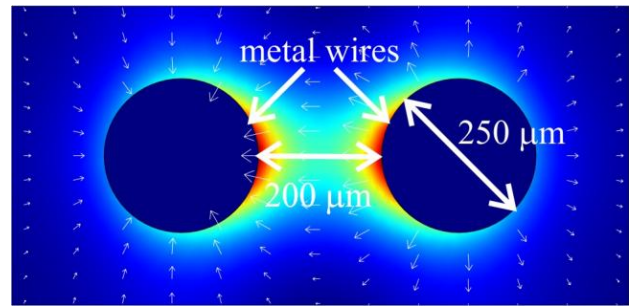

Fig. 1. TEM mode of the metal wire waveguide

In this paper we explore several practical designs of the wire waveguides for THz guidance. The waveguide structure consists of the polyethylene fiber with porous structure and metal wires inserted into it. To fix the wires we need to introduce some dielectric into the waveguide structure. We propose using polyethylene fibers having N+1 openings, where N is the number of the wires used for the waveguide. The central opening remains not occupied by the wire and is used for the guiding of the plasmonic mode of the fibers. High modal confinement in the air reduces the absorption losses. We believe that waveguides described in this paper can be useful not only for low-loss THz wave delivery but also for sensing of biological and chemical specimens in the terahertz region by placing the recognition elements directly into the waveguide microstructure similarly to what has been recently demonstrated in [16].

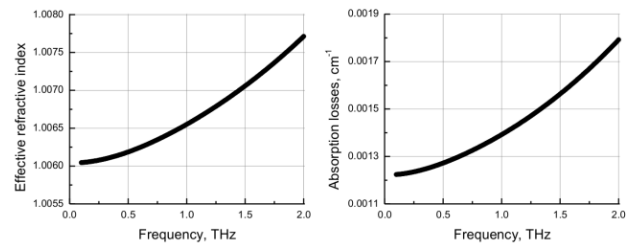

Fig. 2. Effective refractive index and absorption losses of the metal wire waveguide with the parameters depicted in Fig. 1.

Fibers were fabricated using commercial rods of low density polyethylene (LDPE) known to be one of the lowest loss material in the terahertz range. Both wire waveguides preforms were made using a combination of drilling and stacking techniques. Metal wires with the diameter of 250 mm can be either co-drawn or inserted inside the fiber after fabrication (see Fig. 3).

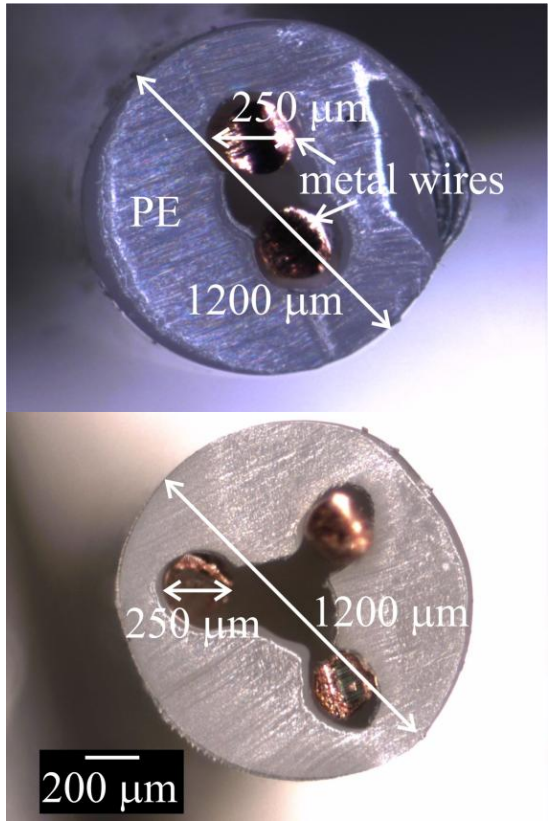

Fig. 3. Photographs of the metal wire waveguide with two and three wires used.

All the data in our experiments was acquired using a modified THz-TDS (Time-Domain Spectroscopy) setup. The setup consists of a frequency-doubled femtosecond fiber laser (MenloSystems C-fiber laser) used as a pump source and identical GaAs dipole antennae used as source and detector yielding a spectrum ranging from ~0.1 to 3 THz. In contrast with a standard arrangement of most THz-TDS setups where the configuration of parabolic mirrors is static, our setup has mirrors mounted on the translation rails. This flexible geometry allows measurement of the waveguides up to 45 cm in length without any realignment.

In Fig. 4 we present experimentally measured electric field at the output of the waveguide. We compare the transmission through the waveguide for the electric fields with two states of polarization, parallel to the wires and perpendicular to the wires.

At low frequencies (< 0.6THz) the only guided mode supported by the waveguide is a total internal reflection

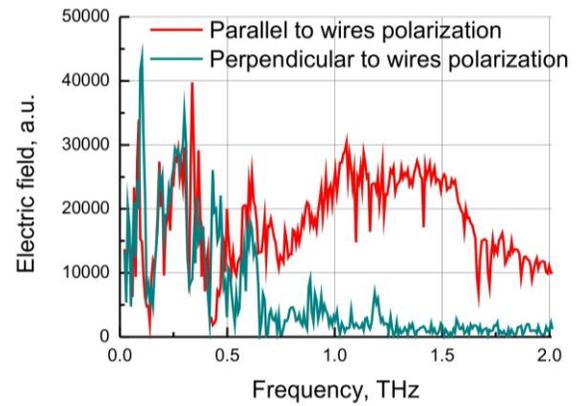

Fig. 4. Experimentally measured electric field in frequency domain at the output of the two-wire waveguide

(TIR) mode where two-wire waveguide acts as a subwavelength waveguide while the fiber diameter is comparable or smaller than the wavelength of light. The wavelength is much higher than the distance between the wires, thus the plasmonic mode is not excited at low frequencies and there is no difference in guiding the radiation with parallel or perpendicular polarization.

At higher frequencies (>0.6THz), the wave guiding is possible only for the light polarized parallel to the wires. The modal pattern represents the mixture of the plasmonic mode of the fibers and the mode of the anti-resonant optical waveguide (ARROW mode) for the parallel polarized radiation (see Fig. 5). Only the ARROW mode guidance is possible for the perpendicularly polarized light with not efficient coupling into it.

To study numerically the propagation in the metal wire waveguides we have imported its cross-section into COMSOL Multiphysics FEM software and then solved for the complex effective refractive indices and field profiles. The attenuation of the two-wire waveguide as a function of frequency also has resonance shape with clearly expressed maxima and minima at higher frequencies (see Fig. 6). The highest achievable decrease of the absorption losses compared to the propagation in bulk material is 40 % at 1.85 THz.

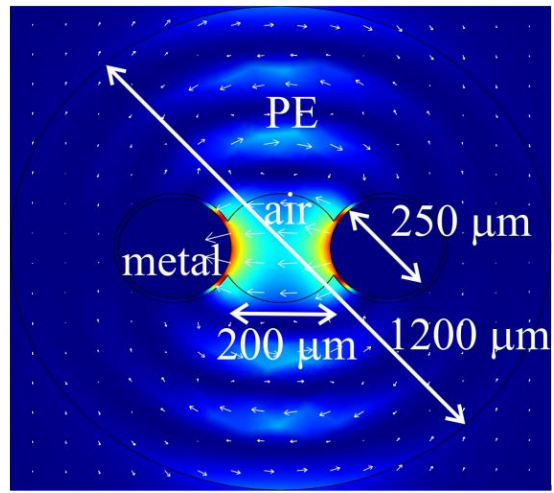

Fig. 5. Mixture of the TEM mode of the metal wires and ARROW mode of the waveguide.

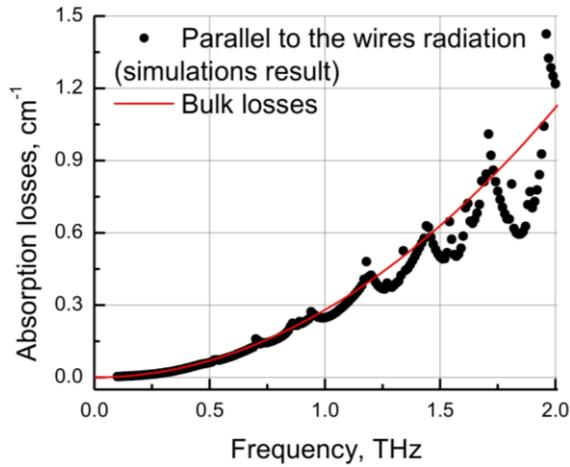

Fig. 6. Result of the numerical simulation of the absorption losses of the metal two-wire waveguide depicted in Fig. 3 (a).

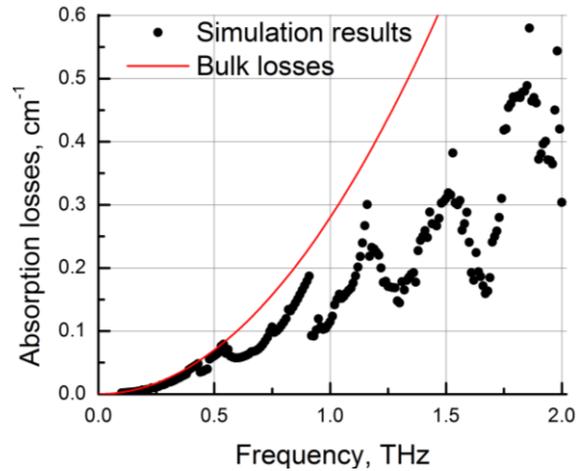

Fig. 8. Result of the numerical simulation of the absorption losses of the metal three-wire waveguide depicted in Fig. 3 (b).

Three-wire waveguide with the larger dimensions of the central air layer compared to the two-wire design has been fabricated and theoretically analyzed. In Fig. 7 we present experimentally measured electric field at the output of the waveguide. The increase of the air part of the waveguide and the distance between the wires leads to the decrease of the attenuation constant and broadens the guidance range of the waveguide, correspondingly.

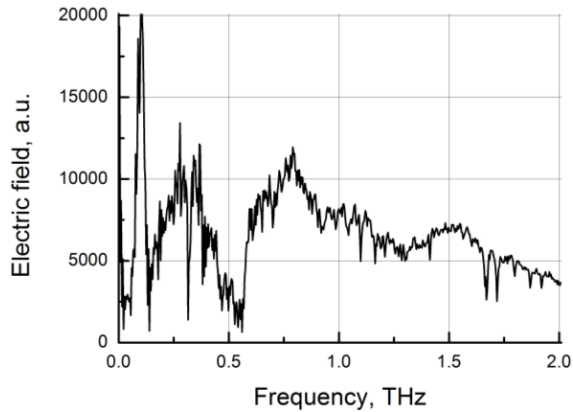

Fig. 7. Experimentally measured electric field in frequency domain at the output of the three-wire waveguide depicted in Fig. 3(b)

As well as in the case of the two-wire waveguide the modal pattern of the three-wire waveguide at frequencies higher than 0.5 THz represents the combination of the plasmonic and ARROW modes. The attenuation of the three-wire waveguide as a function of frequency has resonance shape, the absorption losses can be 3 times lower than in bulk material at certain frequencies (see Fig. 8).

The main difference of the three-wire configuration waveguide over two-wire waveguide is that it is not sensible to the polarization of the initial radiation, making it easier to align the fiber and couple the radiation into it.

We have studied the normalized amplitude coupling coefficients computed from the overlap integral of the respective flux distributions of the waveguide modes with that of the Gaussian beam of the source. The Gaussian beam waist used in the experiments is directly proportional to the operation wavelength. Particularly, in our THz setup we have measured that beam FWHM is about $2.5\lambda$. In Fig. 9 we present the coupling coefficient into the primary mode of the three-wire waveguide. We define the primary mode as the one that has lowest combined coupling and attenuation losses. For frequencies below 0.5 THz the primary mode is the TIR mode and above 0.5 THz the primary mode is the TEM mode of the metal wires affected by the presence of the dielectric fiber.

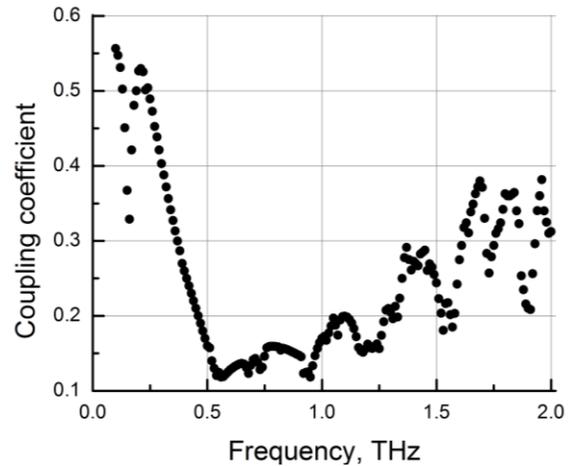

Fig. 9. Coupling coefficient of the three-wire waveguide as a function of frequency. At frequencies < 0.5 THz the guiding mechanism is TIR, above 0.5 THz the waveguide mode represents the mixture of the plasmonic mode of the wires and the ARROW mode of the waveguide.

In conclusion, a novel type of a THz wire waveguide has been proposed combining the low-loss, low group velocity dispersion and simplicity of excitation of the two-wire metal waveguide with the ease of manufacturing, possibility of bending and a convenient access to a modal power of the hollow core waveguide. Optical properties of the wire waveguides are investigated experimentally by THz-TDS spectroscopy and numerically using the finite element method confirming TEM-mode guidance in a broad spectral range.